\documentclass[a4paper,11pt]{article}
\usepackage{pos}
\usepackage{subfigure}

\title{Form factors of the $D_s \to \phi \ell \nu_\ell$ semileptonic decay with (2+1)-flavor lattice QCD}

\author*[a,b]{Gaofeng Fan}
\author[c]{Yu Meng}
\author[d,e,f]{Chuan Liu}
\author[b]{Zhaofeng Liu}
\author[b,g]{Tinghong Shen}
\author[b,h]{Ting-Xiao Wang}
\author[i]{Ke-Long Zhang}
\author[a]{Lei Zhang}

\affiliation[]{\normalsize{\bf \sffamily \hspace{50mm}(The CLQCD Collaboration)}}

\affiliation[a]{School of Physics, Nanjing University, Nanjing 210093, China}
\affiliation[b]{Institute of High Energy Physics, Chinese Academy of Sciences,Beijing 100049, China}
\affiliation[c]{School of Physics, Zhengzhou University, Zhengzhou 450001, China}
\affiliation[d]{School of Physics, Peking University, Beijing 100871, China}
\affiliation[e]{Center for High Energy Physics, Peking University, Beijing 100871, China}
\affiliation[f]{Collaborative Innovation Center of Quantum Matter, Beijing 100871, China}
\affiliation[g]{Hubei Nuclear Solid Physics Key Laboratory, School of Physics and Technology,\\ Wuhan University, Wuhan 430072, China}
\affiliation[h]{School of Physical Sciences, University of Chinese Academy of Sciences, Beijing 101408, China}
\affiliation[i]{Key Laboratory of Quark \& Lepton Physics (MOE) and Institute of Particle Physics,\\ Central China Normal University, Wuhan 430079, China}

\emailAdd{yu\_meng@zzu.edu.cn}
\emailAdd{liuzf@ihep.ac.cn}

\abstract{We present a systematic lattice calculation of the vector and axial vector form factors $V$ and $A_i~(i=0,1,2)$ for the $D_s \to \phi \ell \nu_\ell$ semileptonic decay using (2+1)-flavor Wilson-clover fermion configurations generated by the CLQCD collaboration. Seven gauge ensembles with different lattice spacings, from $0.052~\text{fm}$ to $0.105~\text{fm}$, and different pion masses, from about $210~\text{MeV}$ to $320~\text{MeV}$ are utilized, enabling us to take both the continuum limit and physical pion mass extrapolation. The form factor ratios are obtained to be $r_V=1.614(19)$ and $r_2=0.741(31)$. Our results of form factors reach the precision of $1\%-4\%$, which greatly improves the previous lattice QCD results and obtains the most precise determination to date.}

\FullConference{The 17th International Conference on Heavy Quarks and Leptons (HQL 2025)\\
15-19 September, 2025\\
Peking University, Beijing, China\\}

\begin{document}
\maketitle

\section{Introduction}

Semileptonic decays of $D_s$ mesons provide valuable information to study the weak and strong interactions of charmed hadrons. The partial decay rate is related to the Cabibbo-Kobayashi-Maskawa (CKM) matrix element $|V_{cs}|$, which parametrizes the mixing of different quark flavors, and the form factors describing the non-perturbative strong interactions. Precise calculations of form factors can help to rigorously test the Standard Model and probe potential new physics. Theoretically, the process with a vector final particle is much less studied and poses further challenges due to the decay of the unstable vector particle.

The Euclidean transition matrix element is traditionally parametrized as the following~\cite{Richman:1995wm}
\begin{flalign}
\langle \phi\left(\vec{p}\right)|J_\mu^W\left(0\right)|D_s\left(p^\prime\right)\rangle&=\varepsilon_\nu^*\epsilon_{\mu\nu\alpha\beta} p^{\prime}_{\alpha} p_{\beta}\frac{2V(q^2)}{m+M} +\left(M+m\right)\varepsilon^*_{\mu}A_1(q^2)\nonumber\\
&+\frac{\varepsilon^*\cdot q}{M+m}\left(p+p^\prime\right)_\mu A_2(q^2)-2m\frac{\varepsilon^*\cdot q}{q^2}q_\mu\left[A_0(q^2)-A_3(q^2)\right],
\end{flalign}
where $M$ is the mass of the $D_s$ meson; $m$, $p=(iE,\vec{p})$, and $\varepsilon_\mu$ are the mass, momentum, and polarization vector of the $\phi$ meson; and $q^2=\left(M-E\right)^2-|\vec{p}|^2$ is the transfer momentum square as $D_s$ is at rest. The weak current has the form $J^W_\mu=\bar{s}\gamma_\mu\left(1-\gamma_5\right)c$. Four form factors $V,A_0,A_1,A_2$ are introduced here and  are extracted using the scalar function method, which possesses exact rotational invariance and has been applied to various physical processes~\cite{Tuo:2021ewr,Meng:2021ecs,Meng:2024gpd,Meng:2024nyo,Meng:2024axn}. The scalar functions are constructed by contracting the Lorentzian tensor structure in the transition matrix element. The form factors then can be accessed via solving the linear equations. Then, the key quantities $r_V\equiv V\left(0\right)/A_1\left(0\right)$, $r_2\equiv A_2\left(0\right)/A_1\left(0\right)$, and $r_0\equiv A_0\left(0\right)/A_1\left(0\right)$ can be obtained. The formalism details can be found in Ref.~\cite{Fan:2025qgj}.

\section{Lattice setup and simulation results}

We employ seven $(2+1)$-flavor Wilson-clover gauge
ensembles generated by the CLQCD collaboration~\cite{CLQCD:2023sdb,CLQCD:2024yyn}. The dynamical
ensembles use tadpole-improved tree-level
Symanzik gauge action and tadpole-improved tree-level
clover fermions. The valence strange and charm quark masses are tuned using the “fictitious” meson $\eta_s$ and the $D_s$ meson masses. Parameters of gauge ensembles used in this work can be found in Ref.~\cite{Fan:2025qgj}.

\begin{figure}[htp]
\centering  
\subfigure{
\includegraphics[width=10cm]{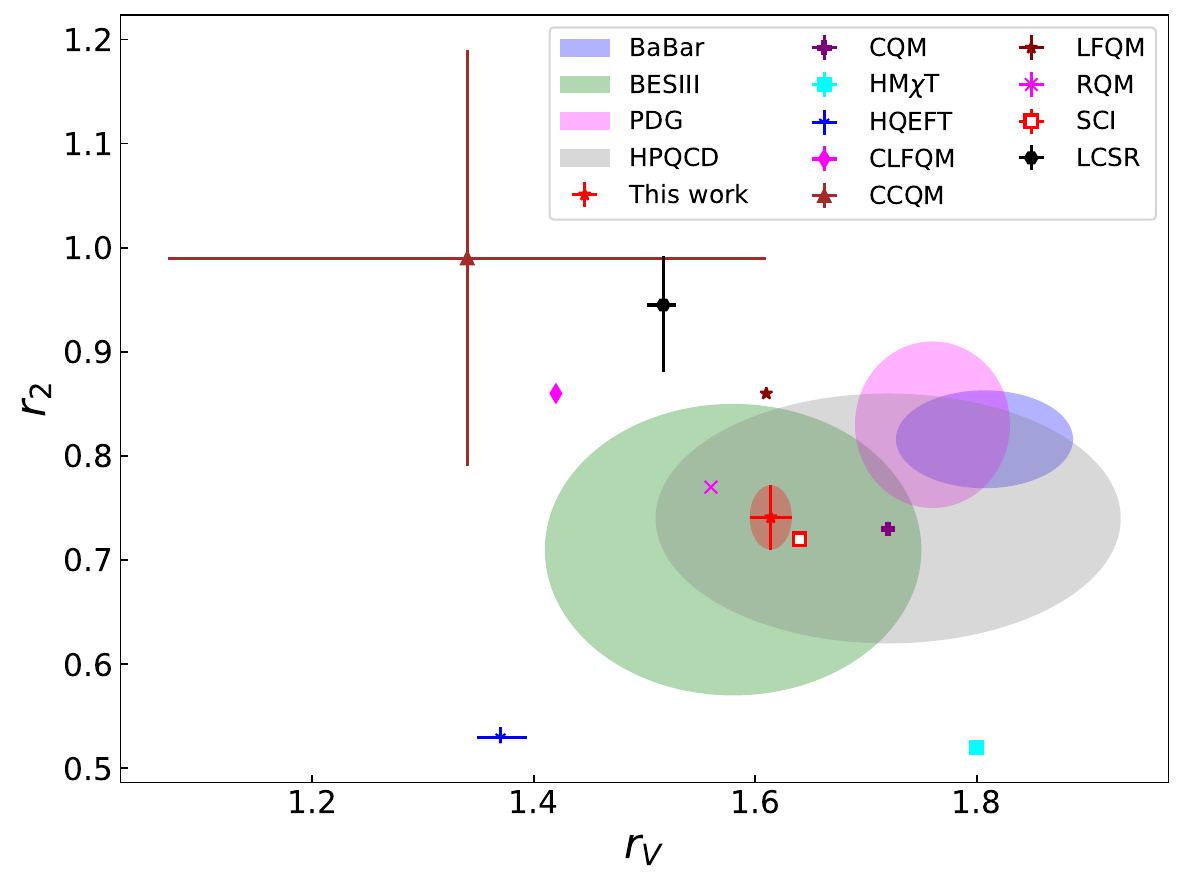}}
\caption{The comparisons of $r_V$ and $r_2$ calculated in this work, and those given by different theoretical predictions~\cite{Melikhov:2000yu,Fajfer:2005ug,Wu:2006rd,Verma:2011yw,Soni:2018adu,Chang:2019mmh,Faustov:2019mqr,Xing:2022sor,Wang:2025oix} and experimental measurements~\cite{BaBar:2008gpr,BESIII:2023opt}.}
\label{rV_r2}
\end{figure}

After obtaining the form factors on each ensemble, the $z$-expansion extrapolation to the continuum limit and physical pion mass point is performed. The fit function is
\begin{flalign}
F\left(q^2,a,m_\pi\right)&=\frac{1}{1-q^2/m_{\text{pole}}^2}\sum_{i=0}^2\left(c_i+d_ia^2\right)\left[1+f_i\left(m_\pi^2-m_{\pi,\text{phys}}^{2}\right)\right]z^{i},
\label{expz}
\end{flalign}
with 
\begin{flalign}
z\left(q^2,t_0\right)=\frac{\sqrt{t_+-q^2}-\sqrt{t_+-t_0}}{\sqrt{t_+-q^2}+\sqrt{t_+-t_0}},
\label{z}
\end{flalign}
where $F$ denotes $V$ and $A_i~(i=0,1,2)$; $t_+=\left(m_{D_s}+m_\phi\right)^2$, $t_0=0$; $c_i$, $d_i$, and $f_i$ are parameters to be determined by fitting. Here, $m_{D_s}$, $m_{\phi}$, and pole masses $m_{D_s^*}$, $m_{D_{s1}}$ for $V$ and $A_i~(i=0,1,2)$ are fixed to their experimental values $m_{D_s}=1968.4~\text{MeV}$, $m_{\phi}=1019.5~\text{MeV}$, $m_{D_s^*}=2112.2~\text{MeV}$ and $m_{D_{s1}}=2459.5~\text{MeV}$~\cite{ParticleDataGroup:2024cfk}. Our results are listed in Table~\ref{limit}. Detailed comparisons of $r_{V,2}$ between our results and previous lattice/phenomenological theory/experiment results are shown in Fig.~\ref{rV_r2}.

\renewcommand{\tablename}{Table}
        \begin{table}[htbp]
	\centering  
     \caption{Numerical results of the form factors $V\left(0\right)$, $A_i\left(0\right)~(i=0,1,2)$, and $r_{V,2,0}$.}
 \label{limit}
 \addvspace{5pt}
	\scalebox{0.75}{\begin{tabular}{ccccccccc}
		\hline\hline\noalign{\smallskip}	
		&$V\left(0\right)$&$A_1\left(0\right)$&$A_2\left(0\right)$&$A_0\left(0\right)$&$A_3\left(0\right)-A_0\left(0\right)$&$r_V$&$r_2$&$r_0$\\
	\noalign{\smallskip}\hline\noalign{\smallskip}
This work &$1.002(9)$&$0.621(5)$&$0.460(19)$&$0.692(4)$&$0.004(12)$&$1.614(19)$&$0.741(31)$&$1.114(11)$\\
HPQCD~\cite{Donald:2013pea}&$1.059(124)$&$0.615(24)$&$0.457(78)$&$0.706(37)$&Enforced to $0$&$1.720(210)$&$0.740(120)$&$1.140(60)$\\
  \noalign{\smallskip}\hline
	\end{tabular}}

\end{table}

\renewcommand{\thesubfigure}{(\roman{subfigure})}
\renewcommand{\figurename}{Figure}


\section{Summary}

In this work, we present a systematic lattice calculation on the $D_s\to\phi\ell\nu_\ell$ semileptonic decay form factors. Seven (2+1)-flavor Wilson-clover gauge ensembles with different lattice spacings and pion masses are used. The calculations cover the full $q^2$ range, leading to a well-controlled accuracy. After the continuum limit and physical pion mass extrapolation, the form factor ratios are obtained to be $r_V=1.614(19)$ and $r_2=0.741(31)$ with $z$-expansion. More results on the branching fractions, $|V_{cs}|$ and discussions on the finite-volume effects can be found in Ref.~\cite{Fan:2025qgj}.

The scalar function scheme employed in this work can be widely applied to other pseudoscalar to vector semileptonic decays, for example, the $D\to K^*$~\cite{BESIII:2024qnx} channel.

\section{Acknowledgments}
The authors acknowledge supports from National Key Research and Development Program of China under Contract No. 2023YFA1606000, and NSFC under Grant No. 12293060, 12293063, 12293065, 12305094, 12192264, 12505099, 12075253. Y.M. also thanks the support from the Young Elite Scientists Sponsorship Program by Henan Association for Science and Technology with Grant No. 2025HYTP003. K.Z. thanks the support from the Cross Research Project of CCNU No. 30101250314. The numerical calculations are supported by the SongShan supercomputer at the National Supercomputing Center in Zhengzhou.

\end{document}